\documentclass[aip,pre,graphicx,reprint]{revtex4-1}
\pdfoutput=1
\usepackage{graphicx}
\usepackage{amssymb}
\usepackage{amsmath}
\usepackage{blkarray}
\usepackage{multirow}
\usepackage{natbib}
\usepackage{epstopdf}

\usepackage{mathpazo}
\usepackage{float}

\begin{document}

\author{C. Perego }

\affiliation{  Department of Chemistry and Applied Biosciences, ETH Zurich, Zurich (Switzerland)} 
\affiliation{  Institute of Computational Science, Universit\`a della Svizzera italiana, Lugano (Switzerland)}

\author{F. Giberti }%
\affiliation{ Institute for Molecular Engineering, The University of Chicago,  Chicago (United States)}

\author{M. Parrinello}
\email{michele.parrinello@phys.chem.ethz.ch}
\affiliation{  Department of Chemistry and Applied Biosciences, ETH Zurich, Zurich (Switzerland)} 
\affiliation{  Institute of Computational Science, Universit\`a della Svizzera italiana, Lugano (Switzerland)}

\title{Chemical Potential Calculations In Dense Liquids Using Metadynamics}

\begin{abstract}
The calculation of chemical potential has traditionally been a challenge in atomistic simulations. One of the most used approaches is Widom's insertion method in which the chemical potential is calculated by periodically attempting to insert an extra particle in the system. In dense systems this method fails since the insertion probability is very low. In this paper we show that in a homogeneous fluid the insertion probability can be increased using metadynamics. We test our method on a supercooled high density binary Lennard-Jones fluid. We find that we can obtain efficiently converged results even when Widom's method fails.
\\
\begin{flushright}
\emph{\small Eur.~Phys.~J.~Spec.~Top.~DOI: 10.1140/epjst/e2016-60094-x} 
\end{flushright}
\end{abstract} 
\maketitle

\section{Introduction}\label{intro}

Chemical potential is an important thermodynamic quantity that regulates many phenomena, such as phase transitions, reaction equilibria and electro-chemistry  \cite{JobEJP2006,BaierleinAJP2001}. The ability to calculate it via a reliable and fast technique is therefore of great value. 
Several numerical methods have been proposed, ranging from Widom test-particle insertion  \cite{WidomJCP1963} to Grand-Canonical Monte-Carlo  approaches  \cite{AdamsMP1974,AdamsMP1975}. Extensive reviews on the efficiency and reliability of these and other techniques can be found in the literature  \cite{KofkeMP1997,ShirtsJCP2005,DalyCPC2012}.
However, despite these efforts, chemical potential calculations still remain a difficult challenge for many real systems, such as dense fluids and solids.

At constant volume and temperature the chemical potential is equal to the Helmoltz Free-Energy difference associated to the addition of an extra molecule. 
The most common approach to calculate this quantity, namely the Widom insertion method, consists in sampling the insertion energy of a test-particle. In this method the largest contribution comes from those configurations in which the insertion of an extra molecule is energetically favorable. This occurs if a cavity sufficiently large to accommodate the extra particle is present. In a low density system many such cavities are present, and Widom method is efficient. On the contrary, in dense systems the formation of a particle-sized cavity is unlikely, and the calculation becomes inefficient.
Thus elaborate methods, based on Voronoi tessellation, have been suggested as a way to look for the appropriate cavities in a dense system \cite{BieshaarMS1995}.
Further suggestions have been among others the introduction of energy-biased sampling of the insertion space  \cite{DelgadoJCP2005}, the use of external potentials to generate controlled inhomogeneities within the system \cite{PowlesJCP1994,MooreJCP2011} or the application of an adaptive resolution simulation scheme \cite{AgarwalJCP2014}. 

In the present paper we shall focus on the calculation of the chemical potential for an homogeneous liquid, and extend the range of densities at which the particle insertion method can be applied by using Well-Tempered (WT) metadynamics  \cite{BarducciPRL2008}. Here the role of metadynamics is to enhance fluctuations in a controlled manner, promoting those rare configurations in which particle insertion is favored.
This leads to an efficient and reliable calculation method, even in those cases in which Widom's method fails.

The paper is organized as follows: in Sec.~\ref{meta} metadynamics main features are recalled, while its application to chemical potential calculation is described in Sec.~\ref{CV}. The implementation of our technique for the selected test case, namely the binary Lennard-Jones (LJ) fluid, is addressed in Sec.~\ref{impl}, while the results of our calculations are reported in Sec.~\ref{results}. Finally the conclusions and perspectives of present work are reported in Sec.~\ref{conclusions}.

\section{Metadynamics}\label{meta}

Before discussing the chemical potential calculation, we recall some of the features of WT metadynamics. For a more detailed description we refer to the by now vast literature on the subject (see e.g.~\onlinecite{ValssonRev2015,BarducciWIR2011}). 

Metadynamics introduces a bias potential to enhance the sampling of rare but important fluctuations. This bias potential is taken to be a function of the so called Collective Variables (CV), a set of order parameters that depend on the atomic coordinates $\mathbf{R}$. The CVs are built to describe the degrees of freedom of the system whose sampling we want to enhance. 
Let us restrict for simplicity to the case of a single collective variable $s(\mathbf{R})$.
The equilibrium distribution of $s$ is given by:
\begin{equation}
\label{pofs}
P(s)=\int\mathrm{d}\mathbf{R}\,\delta[s-s(\mathbf{R})]P(\mathbf{R}),
\end{equation}
where $P(\mathbf{R})=\mathrm{e}^{-\beta U(\mathbf{R})}/Z$ is the Boltzmann distribution defined by potential energy function $U(\mathbf{R})$.  $\beta=(k_{\mathrm{B}}T)^{-1}$ is the inverse temperature and $Z=\int\mathrm{d}\mathbf{R}\mathrm{e}^{-\beta U(\mathbf{R})}$ is the partition function of the system. 
Introducing the free-energy surface $F(s)=-\beta^{-1}\ln P(s)+C$, where $C$ is a constant that plays an inessential role, Eq.~(\ref{pofs}) can be rewritten as
\begin{equation}
\label{pofs2}
P(s)=\frac{\mathrm{e}^{-\beta F(s)}}{\int\mathrm{d}s'\mathrm{e}^{-\beta F(s')}}.
\end{equation}
If a bias potential $V(s)$ is introduced, a different distribution $P_{V}(s)$ is sampled:
\begin{equation}
\label{pvofs}
P_{V}(s)=\int\mathrm{d}\mathbf{R}\,\delta[s-s(\mathbf{R})]P_{V}(\mathbf{R}) = \frac{\mathrm{e}^{-\beta [F(s)+V(s)]}}{\int\mathrm{d}s'\mathrm{e}^{-\beta [F(s')+V(s')]}},
\end{equation}
where $P_{V}(\mathbf{R})=\mathrm{e}^{-\beta [U(\mathbf{R})+V(s(\mathbf{R}))]}/Z_{V}$ is the Boltzmann distribution of the biased ensemble.  $Z_{V}=\int\mathrm{d}\mathbf{R}\mathrm{e}^{-\beta [U(\mathbf{R})+V(s(\mathbf{R}))]}$ is the corresponding partition function.

In WT metadynamics the bias is periodically updated by adding a small repulsive potential, that usually takes the form of a Gaussian of height $w$ and width $\sigma$:
\begin{equation}
\label{Gofs}
G(s,s')=w\mathrm{e}^{-(s-s')^{2}/2\sigma^{2}}.
\end{equation}
$G(s,s')$ is centered on $s'$, that is the value of the CV at the moment of the update. During the simulation the height of the deposited Gaussians is modified in a way that can be described by the following iterative equation:
\begin{equation}
\label{Vn}
V_{n}(s)=V_{n-1}(s)+G(s,s_{n})\exp\left[-\frac{1}{\gamma-1}\beta V_{n-1}(s_{n})\right],
\end{equation}
where $n$ indicates the iteration and $\gamma>1$ is the so called biasing factor, a key parameter in WT metadynamics. After the $n$-th iteration the system evolves under the combined action of $U(\mathbf{R})$ and of $V_{n}(s(\mathbf{R}))$.

It has been shown rigorously \cite{DamaPRL2014} that the bias constructed as in Eq.~(\ref{Vn}) converges asymptotically to:
\begin{equation}
\label{Voft}
V(s)=-\left(1-\frac{1}{\gamma}\right)F(s).
\end{equation}
As a consequence, the biased distribution $P_{V}(s)$ is related to the unbiased one by the relation:
\begin{equation}
\label{reweight}
P_{V}(s)=\frac{\left[P(s)\right]^{1/\gamma}}{\int\mathrm{d}s'\left[P(s')\right]^{1/\gamma} }.
\end{equation}
This result is central to the present paper. In fact Eq.~(\ref{reweight}) shows that narrow peaks in $P(s)$ can be made broader in $P_{V}(s)$ by increasing $\gamma$. Thus rare but important fluctuations, that are hidden in the tails of $P(s)$ are more likely to occur in the biased ensemble.

Furthermore, the theory of metadynamics allows to calculate the unbiased ensemble average $\langle\ldots\rangle$ of any position dependent variable $\mathcal{O}(\mathbf{R})$ via the reweighting \cite{TiwaryJPCB2015,BonomiJCC2009} procedure described by:
\begin{equation}
\label{reweight2}
\langle\mathcal{O}(\mathbf{R})\rangle=\langle \mathcal{O}(\mathbf{R})\mathrm{e}^{\beta\left[V(s(\mathbf{R}),t)-c(t)\right]}\rangle_V,
\end{equation}
where the suffix $V$ indicates that the average is performed under the action of the time varying metadynamics bias potential $V(s(\mathbf{R}),t)$. The position independent constant $c(t)$ is given by \cite{TiwaryJPCB2015}:
\begin{equation}
\label{coft}
c(t)=\beta^{-1}\ln\frac{\int \mathrm{d}s\,\exp\left[\frac{\gamma}{\gamma-1}\beta V(s,t)\right]}{\int \mathrm{d}s\,\exp\left[\frac{1}{\gamma-1}\beta V(s,t)\right]}.
\end{equation}

\section{Choice of the Collective Variable}\label{CV}
We now define the proper CV that will be used to calculate the chemical potential via metadynamics.
Let us begin by considering a system of $N$ particles with coordinate vector $\mathbf{R}=(\mathbf{R}_{1},\ldots,\mathbf{R}_{N})$. As pointed out by Widom \cite{WidomJCP1963}, the excess chemical potential, that is the chemical potential from which the free particle contribution has been subtracted, can be written as:
\begin{equation}
\label{widom}
\mu^{\mathrm{ex}}=-\beta^{-1}\ln\left\langle\frac{1}{\mathcal{V}}\int\exp\left[-\beta \Delta U(\mathbf{R}^{*};\mathbf{R})\right]\mathrm{d}\mathbf{R}^{*}\right\rangle,
\end{equation}
where $\mathcal{V}$ is the system volume and
\begin{equation}
\label{DU}
\Delta U(\mathbf{R}^{*};\mathbf{R})=U(\mathbf{R}^{*};\mathbf{R})-U(\mathbf{R}),
\end{equation}
where $U(\mathbf{R}^{*};\mathbf{R})$ is the potential energy of the system with an extra particle in $\mathbf{R}^{*}$. The average $\langle\ldots\rangle$ in Eq.~(\ref{widom}) is performed over the Boltzmann distribution defined by the $N$ particle interaction potential $U(\mathbf{R})$.
Widom suggested to estimate $\mu^{\mathrm{ex}}$ by evolving the system and periodically computing $\Delta U(\mathbf{R}^{*};\mathbf{R})$ at a randomly generated set of positions $\mathbf{R}^{*}$. Also in other methods, first the coordinate vector $\mathbf{R}$ is sampled and later insertions are attempted \cite{BieshaarMS1995,BennettJCP1976}.

Here we take an alternative route and first we fix a number of insertion points $\mathbf{R}_{i}^{*}$ and later we let $\mathbf{R}$ adjust around them so that the sampling of $\Delta U(\mathbf{R}_{i}^{*};\mathbf{R})$ is enhanced. In order to see how this is done we restrict to the case of an homogeneous fluid, in which the average $\langle\exp\left[-\beta \Delta U(\mathbf{R}^{*};\mathbf{R})\right]\rangle$ is $\mathbf{R}^{*}$ independent. It follows that one can choose a set of $M$ $\mathbf{R}^{*}_{i}$'s and rewrite the chemical potential as:
\begin{equation}
\label{widom2}
\mu^{\mathrm{ex}}=-\beta^{-1}\ln\left\langle\frac{1}{M}\sum_{i=1}^{M}\exp\left[-\beta \Delta U(\mathbf{R}^{*}_{i};\mathbf{R})\right]\right\rangle.
\end{equation}
This form of $\mu^{\mathrm{ex}}$ suggests the introduction of the following CV:
\begin{equation}
\label{sform}
s(\mathbf{R})=-\beta^{-1}\ln\left(\frac{1}{M}\sum_{i=1}^{M}\exp\left[-\beta\Delta U(\mathbf{R}^{*}_{i};\mathbf{R})\right]\right).
\end{equation}
In principle even the choice $M=1$, namely that of using a single insertion point $R^{*}$, leads to the correct result, however it requires much longer sampling times. In this paper the insertion points $R_{i}^{*}$ will be arranged on a regular grid, but other choices are possible.

A consequence of this CV definition is that the chemical potential can be rewritten as the expectation value of a function of $s$:
\begin{equation}
\label{muofs}
\mu^{\mathrm{ex}}=-\beta^{-1}\ln\left\langle\mathrm{e}^{-\beta s}\right\rangle,
\end{equation}
or, in terms of the free energy associated to $s$, as:
\begin{equation}
\label{muofs2}
\mu^{\mathrm{ex}}=-\beta^{-1}\ln\frac{\int \mathrm{e}^{-\beta [F(s)+s]}\mathrm{d}s}{\int\mathrm{e}^{-\beta F(s)}\mathrm{d}s}.
\end{equation}
From Eqs.~(\ref{muofs}) or (\ref{muofs2}) it follows that the largest contribution to $\mu^{\mathrm{ex}}$ comes from those configurations for which $s$ has a negative value. 

This choice of CV has a practical drawback. In fact in most models the atoms interact at short distances via a highly repulsive and rapidly varying potential. Thus, if any of the insertion points $\mathbf{R}_{i}^{*}$ comes close to the atomic positions, $s(\mathbf{R})$ varies very rapidly and can even diverge. This can be remedied by modifying the insertion potential at short distances using a regularized potential $U^{\mathrm{r}}$ in the definition of a new and more smoothly behaved CV:
\begin{equation}
\label{srform}
s^{\mathrm{r}}(\mathbf{R})=-\beta^{-1}\ln\left(\frac{1}{M}\sum_{i=1}^{M}\exp\left[-\beta\Delta U^{\mathrm{r}}(\mathbf{R}^{*}_{i};\mathbf{R})\right]\right).
\end{equation}
The idea is that $\Delta U^{\mathrm{r}}(\mathbf{R}^{*}_{i};\mathbf{R})$ should be large but finite when $\mathbf{R}^{*}$ is close to one of the atomic positions, so that the derivatives $\partial s^{\mathrm{r}}/\partial\mathbf{R}$ are of manageable size.
One can then use $s^{\mathrm{r}}$ to bias the metadynamics simulation and calculate $F(s)$ using the reweighting procedure in Eq.~(\ref{reweight2}) \cite{TiwaryJPCB2015}. 
However, in the test case that will be discussed below, we define $U^{\mathrm{r}}$ so that the change of CV has no practical effect on the calculation of $\mu^{\mathrm{ex}}$, and the reweighting is not necessary (see also the Supporting Information (SI)).

\section{Test Case}\label{impl}

In order to test the method we shall consider a binary Lennard-Jones (LJ) fluid. In particular we consider the system studied by Kob and Andersen \cite{KobPRL1994,KobPRE1995} for its glass forming properties, that it is known to remain fluid at low temperature and high density. Thus it provides a challenging test for our method. The system is composed by two atomic species, $A$ and $B$, of equal atomic mass. The interaction potential is:
\begin{equation}
\label{Ubin}
U_{\alpha\alpha'}(r)=4\epsilon_{\alpha\alpha'}\left[\left(\frac{\sigma_{\alpha\alpha'}}{r}\right)^{12}-\left(\frac{\sigma_{\alpha\alpha'}}{r}\right)^{6}\right],
\end{equation}
where $r$ is the atom-atom distance and $\alpha,\alpha'= {A,B}$. In such a system one defines an excess chemical potential $\mu^{\mathrm{ex}}_{\alpha}$ for each species, and generalizes Eq.~(\ref{widom2}):
\begin{equation}
\label{widombin}
\mu^{\mathrm{ex}}_{\alpha}=-\beta^{-1}\ln\left\langle\frac{1}{M}\sum_{i=1}^{M}\exp\left[-\beta \Delta U_{\alpha}(\mathbf{R}_{i}^{*};\mathbf{R})\right]\right\rangle,\ \alpha={A,B}.
\end{equation}
where
\begin{equation}
\label{DUbin}
\Delta U_{\alpha}(\mathbf{R}^{*};\mathbf{R})=\sum_{\alpha'=A,B}\sum_{i=1}^{N^{\alpha'}}U_{\alpha\alpha'}(\mathbf{R}^{*}-\mathbf{R}^{\alpha'}_{i}),
\end{equation}
where $\mathbf{R}^{A}_{i}$ and $\mathbf{R}^{B}_{i}$ are the coordinates of the $A,B$ atoms and $\mathbf{R}$ indicates the ensemble of the two combined sets of coordinates. $N^{\alpha'}$ is the number of $\alpha'$ specie particles.

Each species will have its own $s$, therefore Eq.~(\ref{sform}) generalizes to:
\begin{equation}
\label{sbin}
s_{\alpha}=-\beta^{-1}\ln\left(\frac{1}{M}\sum_{i=1}^{M}\exp\left[-\beta\Delta U_{\alpha}(\mathbf{R}^{*}_i;\mathbf{R})\right]\right),\ \alpha={A,B}.
\end{equation}
Analogously, Eq.~(\ref{muofs}) becomes:
\begin{equation}
\label{mubins}
\mu^{\mathrm{ex}}_{\alpha}=-\beta^{-1}\ln\left\langle\mathrm{e}^{-\beta s_{\alpha}}\right\rangle,\ \alpha={A,B},
\end{equation}
and Eq.~(\ref{muofs2}) becomes:
\begin{equation}
\label{muofsbin}
\mu^{\mathrm{ex}}_{\alpha}=-\beta^{-1}\ln\frac{\int \mathrm{e}^{-\beta [F(s_{\alpha})+s_{\alpha}]}\mathrm{d}s_{\alpha}}{\int\mathrm{e}^{-\beta F(s_{\alpha})}\mathrm{d}s_{\alpha}},\ \alpha={A,B}.
\end{equation}
As mentioned before, in order to avoid divergences and the appearance of a large bias, we shall use the regularized version $s^{\mathrm{r}}_{\alpha}$ of the CV, defining the following modified potential:
\begin{equation}
\label{Umod}
U^{\mathrm{r}}_{\alpha\alpha'}(r)=
\begin{cases}
\frac{1}{2 r_{0}}\frac{\partial U_{\alpha\alpha'}}{\partial r}\Big|_{r=r_{0}}\left(r^{2}-r_{0}^{2}\right)+U_{\alpha\alpha'}(r_{0}) & \mbox{if } r\leq r_{0} \\
U_{\alpha\alpha'}(r) & \mbox{if } r > r_{\mathrm{0}}\\
\end{cases},
\end{equation}
where $r_{0}$ is the distance below which $U^{\mathrm{r}}_{\alpha\alpha'}$ and  $U_{\alpha\alpha'}$ differ. 

\section{Results and Discussion}\label{results}
\begin{figure*}[ht] 
\includegraphics[scale=0.42]{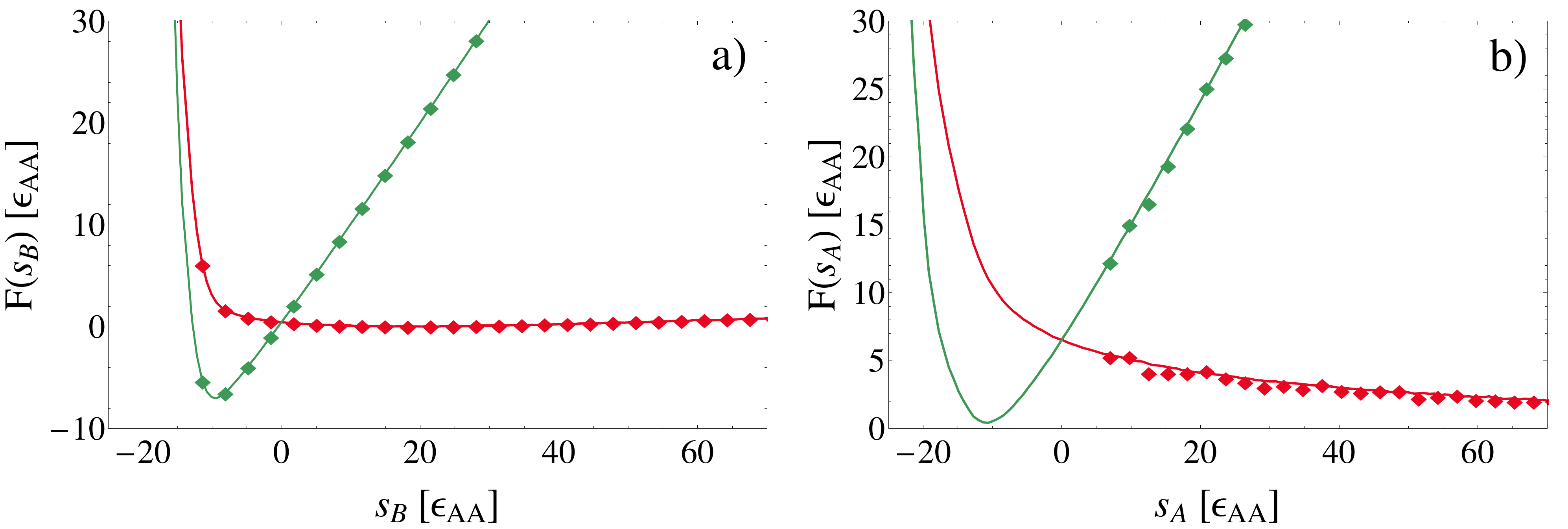}\\  
\caption{\label{FsAB} Plot of $F(s_{\alpha})$ (red) and  $F(s_{\alpha})+s_{\alpha}$ (green) for $\alpha=B$ (panel a) and $\alpha=A$ (panel b). The diamonds display the results of the unbiased simulation, in which the free energy is computed from the histogram of $s_{\alpha}$ with $M=64$ (for clarity of presentation we have plotted only a subset of values). The lines display the results of the WT metadynamics simulations, in which $s^{\mathrm{r}}_{\alpha}$ is biased and the resulting distribution is reweighted to obtain $F(s_{\alpha})$. In the SI we show that $F(s^{\mathrm{r}}_{\alpha})$ and $F(s_{\alpha})$ are equal in the relevant region for the computation of $\mu^{\mathrm{ex}}_{\alpha}$. The reported free energy surfaces are the result of $5\times10^{5}\,\Delta\tau$ long runs. }\end{figure*}
Kob and Andersen model \cite{KobPRE1995} is a mixture of $80\,\%$ $A$ particles and $20\,\%$ $B$ particles. In units of $\sigma_{AA}$ and $\epsilon_{AA}$ the other potential parameters are $\sigma_{BB}=0.88$, $\epsilon_{BB}=0.5$, $\sigma_{AB}=0.8$ and $\epsilon_{AB}=1.5$. In our simulations the potentials $U_{\alpha\alpha'}$ are cut and shifted at $r_{\mathrm{c}}=2.5$, while the distance parameter of the regularized potential $U^{\mathrm{r}}_{\alpha\alpha'}$ is set at $r_{0}=0.88\, \sigma_{\alpha\alpha'}$. 
The simulated system is composed of $N=800$ atoms in a cubic box of length $L=8.73$ and Periodic Boundary Conditions (PBC) are applied.
The temperature is kept at $k_{\mathrm{B}}T=3/4$ by using the stochastic velocity rescaling thermostat  \cite{BussiJCP2007}. At these thermodynamic conditions the system is in the liquid phase.

The MD simulations were carried out using the GROMACS \cite{HessJCTC2008} software, equipped with a private version of the PLUMED2 plug-in  \cite{TribelloCPC2014}. The integration time-step was chosen to be $\Delta t= 9.3\times 10^{-4}$ (which corresponds to $\Delta t= 2\,\mathrm{fs}$ for the LJ parameters of Argon: $M=39.948\,\mathrm{u}$, $\epsilon_{AA}=0.996\,\mathrm{kJ/mol}$ and $\sigma_{AA}=3.405$ \AA). The Widom's calculations were performed using the appropriate GROMACS subroutines. 

One advantage of using this particular binary system is that the majority of atoms ($A$) have a larger radius, leaving sufficient interstices to place the smaller $B$ atoms. This makes the calculation of $\mu_{B}^{\mathrm{ex}}$ relatively easy and accessible to Widom's method, providing a testing ground for our technique. On the other hand the calculation of $\mu_{A}^{\mathrm{ex}}$ is challenging and its successful completion demonstrates the usefulness of our approach.
We performed two separate calculations, for $\mu^{\mathrm{ex}}_{A}$ and $\mu^{\mathrm{ex}}_{B}$. The Gaussian initial height was chosen to be of $1.2$ and the width is equal to $1.0$. New Gaussians were deposited every $\Delta\tau=500\,\Delta t$ and the $\gamma$ parameter was $15$.

We experimented with different values of number of insertion points $M$ (see Eq.~(\ref{sbin})). They all lead to the same results within statistical error. However, the speed of convergence of $\mu^{\mathrm{ex}}$ was different, with the limiting $M=1$ case being particularly slow (see the SI).
Here we present the results for a regular mesh of $M=4\times4\times4=64$ points. While $M$ was not carefully optimized, this choice seemed to yield a good compromise between speed of convergence and computational cost.
We note that the calculations can be easily parallelized and it pays to distribute wisely the load among the concurrent processors, thus the determination of the optimal $M$ depends also on the computational platform.

As discussed earlier we expect the chemical potential calculation to be easier for the smaller atoms ($B$) and more challenging for the large component ($A$). This is brought out in Fig.~\ref{FsAB}, where the $F(s_{\alpha})$ resulting from metadynamics simulations is compared to the $F(s_{\alpha})$ calculated in an unbiased MD run. In the $B$ case (Fig.~\ref{FsAB}a), even without applying the bias potential, the range of spontaneously occurring fluctuations, indicated in figure by the domain of the unbiased $F(s_{\alpha})$, is large enough to overlap with the negative $s_{B}$ region. This region yields the more substantial contribution to the numerator of Eq.~(\ref{muofsbin}), and thus to the $\mu^{\mathrm{ex}}_{B}$ value (see SI for further details). Thus $\mu^{\mathrm{ex}}_{B}$ can be reliably computed with unbiased sampling. Note however that even in this favorable case the spontaneous fluctuations do not cover entirely the tail of the distribution. As noted in Ref.~\cite{LuJCP2001a} this leads to a systematic error. In contrast, no such problem is present in the metadynamics calculation, where the fluctuations of $s_{B}$ cover thoroughly all the important region.

\begin{figure*}[ht] 
\includegraphics[scale=0.40]{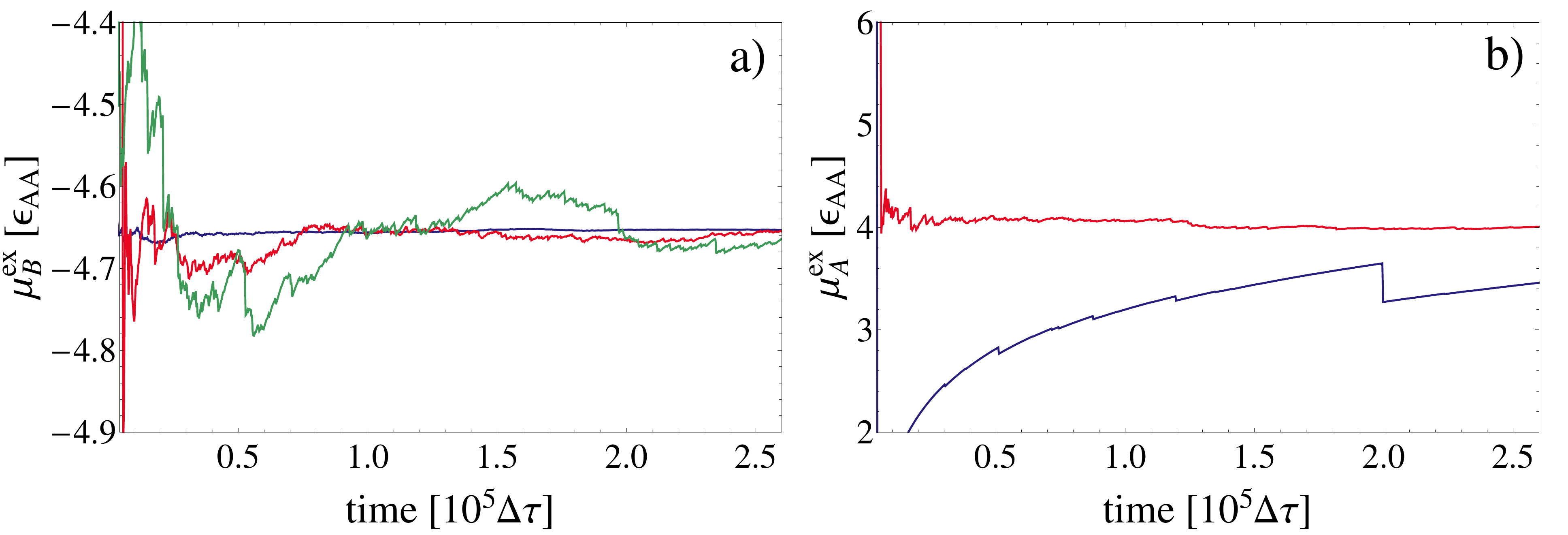}\\  
\caption{\label{muAB} Plot of $\mu^{\mathrm{ex}}_{\alpha}$ as a function of MD steps. In the $\alpha=B$ case (panel a), we compare the results of the metadynamics run  with $M=64$ (red line) to a Widom calculation in which $64$ insertions are attempted every $\Delta\tau$ (green line). As a reference we report also the result of a more accurate Widom calculation with $5\times10^{4}$ insertions per $\Delta\tau$ (blue line). In the $\alpha=A$ case (panel b), we compare the results of the metadynamics run  with $M=64$ (red line) to a Widom calculation using $10^{6}$ insertions per $\Delta\tau$ (blue line). 
}\end{figure*}

In the case of the larger atoms (Fig.~\ref{FsAB}b) the situation is much more dramatic and the unbiased results show little or no overlap between the fluctuations of $s_{A}$ and the region that contributes the most to $\mu^{\mathrm{ex}}_{A}$. In contrast, our metadynamics calculations explore in detail the important region of $F(s_{A})$. This is reflected by the convergence of $\mu^{\mathrm{ex}}_{A}$ and $\mu^{\mathrm{ex}}_{B}$ calculations, shown in Fig.~\ref{muAB}, where we compare metadynamics results (obtained with Eq.~(\ref{reweight2})) to standard Widom calculations. In the $B$ case (Fig.~\ref{muAB}a) the agreement between metadynamics and standard Widom method is within the statistical error. In the $A$ case however (Fig.~\ref{muAB}b) Widom's method faces severe difficulties and convergence is jumpy and very slow, even with a very large number of insertion points sampled. On the contrary our method converges to an accuracy of $1\,\%$ using only $M=4\times4\times4=64$ insertion points. This underlines the necessity of metadynamics to visit the extreme tails of the $s$ distribution and calculate $\mu^{\mathrm{ex}}$ accurately.

\section{Conclusions}\label{conclusions}
In conclusion we have shown that this new approach can be very competitive with previous ones. Three features should be noted: the use of the homogeneity of the system, the change of paradigm from an atom position-centric to an insertion-centric point of view and the use of metadynamics to enhance fluctuations. 
The calculation is no more difficult than a standard WT metadynamics one, and can be performed using any of the codes with which the PLUMED2 plug-in can be integrated. 
The recently developed variationally enhanced sampling \cite{ValssonPRL2014}, in its WT version \cite{ValssonJCTC2015}, could be also profitably applied to perform these calculations and lead to further improvement in the efficiency of chemical potential evaluations. 
Although here our method has been developed for homogeneous systems, its extension to non-homogeneous situations is possible and it is currently under study.
A relatively straightforward application of this method is the calculation of the solvation energy of molecules in liquids, another quantity of great practical interest. Moreover, if properly adapted, these ideas could be extended to free energy perturbation calculations  \cite{ZwanzigJCP1954,Allenbook1989,KollmanCR1993,Frenkelbook2001}.

\begin{acknowledgements}
The authors are pleased to acknowledge M.~Nava, B.~Smit and O.~Valsson for the useful discussions. The computational resources were provided by the Swiss Center for Scientific Computing, and the Brutus Cluster at ETH Zurich. The authors acknowledge also the VARMET European Union Grant ERC-2014-ADG-670227 and the National Centres of Competence in Research ``Materials Revolution: Computational Design and Discovery of Novel Materials'' project for funding.

\end{acknowledgements}

%

%
%



\begin{thebibliography}{10}

\bibitem{JobEJP2006}
G.~Job, F.~Herrmann, Eur.~J.Phys.~\textbf{27}(2), (2006) 353

\bibitem{BaierleinAJP2001}
R.~Baierlein, Am.~J.~Phys.~\textbf{69}(4), (2001) 423 

\bibitem{WidomJCP1963} 
B.~Widom, J.~Chem.~Phys.~\textbf{39}(11), (1963) 2808

\bibitem{AdamsMP1974}
D.~J.~Adams, Mol.~Phys.~\textbf{28}(5), (1974) 1241

\bibitem{AdamsMP1975}
D.~J.~Adams, Mol.~Phys.~\textbf{29}(1), (1975) 307

\bibitem{KofkeMP1997}
D.~A.~Kofke, P.~T.~Cummings, Mol.~Phys.~\textbf{92}(6), (1997) 973

\bibitem{ShirtsJCP2005}
M.~R.~Shirts, V.~S.~Pande, J.~Chem.~Phys.~\textbf{122}(14), (2005) 144107

\bibitem{DalyCPC2012}
K.~B.~Daly, J.~B.~Benziger, P.~G.~Debenedetti, A.~Z.~Panagiotopoulos, Comput.~Phys.~Commun.~\textbf{183}(10), (2012) 2054

\bibitem{BieshaarMS1995}
R.~Bieshaar, A.~Geiger, N.~N.~Medvedev, Mol.~Simulat.~\textbf{15}(3), (1995) 189

\bibitem{DelgadoJCP2005}
R.~Delgado-Buscalioni, G.~De~Fabritiis, P.~V.~Coveney, J.~Chem.~Phys.~\textbf{123}(5), (2005) 054105

\bibitem{PowlesJCP1994}
J.~G.~Powles, B.~Holtz, W.~A.~B.~Evans, J.~Chem.~Phys.~\textbf{101}(9), (1994) 7804

\bibitem{MooreJCP2011}
S.~G.~Moore, D.~R.~Wheeler, J.~Chem.~Phys.~\textbf{134}(11), (2011) 114514

\bibitem{AgarwalJCP2014}
A.~Agarwal, H.~Wang, C.~Sch\"utte, L.~Delle Site, J.~Chem.~Phys.~\textbf{141}(3), (2014) 034102

\bibitem{BarducciPRL2008}
A.~Barducci, G.~Bussi, M.~Parrinello, Phys.~Rev.~Lett.~\textbf{100}(2), (2008) 020603

\bibitem{ValssonRev2015}
O.~Valsson, P.~Tiwary, and M.~Parrinello, Annu.~Rev.~Phys.~Chem.~\textbf{67}(1), (2016) 

\bibitem{BarducciWIR2011}
A.~Barducci, M.~Bonomi, M.~Parrinello, Wiley Interdiscip.~Rev.~Comput.~Mol.~Sci.~\textbf{1}(5), (2011) 826

\bibitem{DamaPRL2014}
J.~F.~Dama, M.~Parrinello, G.~A.~Voth, Phys.~Rev.~Lett.~\textbf{112}(24), (2014) 240602

\bibitem{TiwaryJPCB2015}
P.~Tiwary, M.~Parrinello, J.~Phys.~Chem.~B \textbf{119}(3), (2015) 736 

\bibitem{BonomiJCC2009}
M.~Bonomi, A.~Barducci, M.~Parrinello, J.~Comput.~Chem.~\textbf{30}(11), (2009) 1615

\bibitem{BennettJCP1976}
C.~H.~Bennett, J.~Comput.~Phys.~\textbf{22}(2), (1976) 245

\bibitem{KobPRL1994}
W.~Kob, H.~C.~Andersen, Phys.~Rev.~Lett.~\textbf{73}(10), (1994) 1376

\bibitem{KobPRE1995}
W.~Kob, H.~C.~Andersen, Phys.~Rev.~E \textbf{51}(5), (1995) 4626

\bibitem{BussiJCP2007}
G.~Bussi, D.~Donadio, M.~Parrinello, J.~Chem.~Phys.~\textbf{126}(1), (2007) 014101

\bibitem{HessJCTC2008}
B.~Hess, C.~Kutzner, D.~van der Spoel, E.~Lindahl, J.~Chem.~Theory Comput.~\textbf{4}(3), (2008) 435

\bibitem{TribelloCPC2014}
G.~A.~Tribello, M.~Bonomi, D.~Branduardi, C.~Camilloni, G.~Bussi, Comput.~Phys.~Commun.~\textbf{185}(2), (2014) 604

\bibitem{LuJCP2001a}
N.~Lu, D.~A.~Kofke, J.~Chem.~Phys.~\textbf{114}(17), (2001) 7303

\bibitem{ValssonPRL2014}
O.~Valsson, M.~Parrinello, Phys.~Rev.~Lett.~\textbf{113}(9), (2014) 090601

\bibitem{ValssonJCTC2015}
O.~Valsson, M.~Parrinello, J.~Chem.~Theory Comput.~\textbf{11}(5), (2015) 1996

\bibitem{ZwanzigJCP1954}
R.~W.~Zwanzig, J.~Chem.~Phys.~\textbf{22}(8), (1954) 1420

\bibitem{Allenbook1989}
M.~P.~Allen, D.~J.~Tildesley, \textit{Computer simulation of liquids} (Oxford university press, 1989)

\bibitem{KollmanCR1993}
P.~Kollman, Chem.~Rev.~\textbf{93}(7), (1993) 2395

\bibitem{Frenkelbook2001}
D.~Frenkel, B.~Smit, \textit{Understanding molecular simulation: from algorithms to applications} (Academic press, 2001)

\end{thebibliography}

\end{document}